\documentclass{revtex4}

\usepackage{amssymb}
\usepackage{amsfonts}
\usepackage{amstext}
\usepackage{graphicx}
\usepackage{amscd}
\usepackage{amsmath}

\newcommand{\ihbar}{\imath \hbar}

\newcommand{\Te}{\mathbb{T}e}
\newcommand{\T}{\mathbb T}

\renewcommand{\P}{\mathbb{P}}
\newcommand{\hol}{\mathrm{Hol}}

\newtheorem{defi}{Definition}
\newtheorem{prop}{Property}
\newtheorem{propo}{Proposition}
\newtheorem{theo}{Theorem}

\begin{document}

\title{Holonomy of a principal composite bundle connection, non-abelian geometric phases and gauge theory of gravity}

\author{David Viennot}
\email{david.viennot@utinam.cnrs.fr}
\affiliation{Institut UTINAM (CNRS UMR 6213, Universit\'e de Franche-Comt\'e, Observatoire de Besan\c con), 41bis Avenue
  de l'Observatoire, BP1615, 25010 Besan\c con cedex, France.}
 
\begin{abstract}
We show that the holonomy of a connection defined on a principal composite bundle is related by a non-abelian Stokes theorem to the composition of the holonomies associated with the connections of the component bundles of the composite. We apply this formalism to describe the non-abelian geometric phase (when the geometric phase generator does not commute with the dynamical phase generator). We find then an assumption to obtain a new kind of separation between the dynamical and the geometric phases. We also apply this formalism to the gauge theory of gravity in the presence of a Dirac spinor field in order to decompose the holonomy of the Lorentz connection into holonomies of the linear connection and of the Cartan connection.
\end{abstract}

\maketitle

\section{Introduction}
A composite bundle is a tower of bundles $E(T,F_2) \to T(B,F_1) \to B$ (the notation $T(B,F)\xrightarrow{\pi} B$ denotes a locally trivial fibre bundle, with base space $B$, typical fibre $F$, total space $T$ and projection $\pi$, with $B$, $F$ and $T$ being three $\mathcal C^\infty$-manifolds and $\pi$ being a surjective map). The concept of composite bundle was introduced by Sardanashvily in ref. \cite{sardana}, with $E(T,F_2) \to T$ a vector bundle, to describe both the non-abelian geometric phase and the non-abelian dynamical phase. Sardanashvily ref. \cite{sardana2,sardana3} and Tresguerres ref. \cite{tresguerres} have also used composite bundles to describe the gauge theory of gravity. To complete the description of a non-abelian geometric phase with a non-abelian dynamical phase, we have introduced in ref. \cite{viennot} the concept of principal composite bundle. A principal composite $P(M,G)$-bundle $P_+(S,G) \to S(R,M) \to M$ mimes a principal bundle, where a base fibre bundle $S(R,M) \to R$ plays the role of the base manifold, and where a structure principal $G$-bundle $P(M,G) \to M$ plays the role of the structure group ($R$, $M$, $S$, $P$ and $P_+$ are $\mathcal C^\infty$-manifolds, and $G$ is a Lie group). We have shown that a principal composite bundle defines a locally defined $G$-bundle $P_+(M\times R,G) \to M \times R$ called total bundle, and two kinds of ``leaf'' of fibres. One concists of bundles isomorphic to $P(M,G) \to M$, and the others are called transversal $G$-bundles ($Q_x(N,G) \to N$). In the present paper, we specify the principal composite bundle structure, particularly from the viewpoint of the local data and of the torsion of the total bundle. Moreover we study the holonomy of a composite connection and we show that it is related by a non-abelian Stokes theorem to the product of the holonomy of the structure bundle connection by the holonomy of one of the transversal bundle connections.\\
This paper is organized as follows. Section II presents the principal composite bundle geometry. Section III studies composite connections and the associated holonomies. The link between composite holonomies and geometric phases, when the geometric phase generator does not commute with the dynamical phase generator, is studied in section IV. Under a relevant assumption we find that the non-abelian phase is the product of the non-abelian geometric phase and of the non-abelian dynamical phase with fixed adiabatic parameters. Finally, we consider the composite holonomies arising with a Dirac spinor field transport in a curved spacetime.\\
{\it A note about the notations used here : the symbol ``$\simeq$'' between two manifolds denotes that the two manifolds are diffeomorphic. The symbol ``$\hookrightarrow$'' denotes  an inclusion between two sets. $\Omega^n(M,\mathfrak g)$ denotes the set of the $\mathfrak g$-valued $n$-forms of the manifold $M$. $d_M$ denotes the exterior differential of $M$. $T_xM$ denotes the space of tangent vectors of $M$ at $x \in M$ and $TM$ denotes the space of tangent vector fields of $M$ (the tangent bundle of $M$). $\mathcal L_{x}M$ denotes the space of smooth closed loops in $M$ with base point $x$. $f \circ g(x)$ denotes the composition $f(g(x))$. $\Gamma(U,P)$ denotes the set of local sections of a fiber bundle $P \to M$ over $U \subset M$.}

\section{The principal composite bundle geometry}
This section presents the relevent definitions concerning the principal composite bundles. The first part presents the global theory (as in ref. \cite{viennot}) but with enlightenment concering the torsion of the total bundle; the second part introduces the local theory (which is not treated in ref. \cite{viennot}).
\subsection{Global geometry of a principal composite bundle}
\begin{defi}[Principal composite bundle]
Let $G$ be a Lie group and $P(M,G) \xrightarrow{\pi_P} M$ be a principal $G$-bundle over a manifold $M$. A principal composite $P(M,G)$-bundle, is a tower of fibre bundles $P_+(S,G) \xrightarrow{\pi_+} S(R,M) \xrightarrow{\pi_S} R$ such that
\begin{itemize}
\item $\forall y \in R$, $\pi^{-1}_S(y) \simeq M$,
\item for ervery good open cover $\{V^i\}_i$ of $R$, there exists a diffeomorphism $\chi^{i}_S : \pi_S^{-1}(V^i) \xrightarrow{\simeq} V^i \times M$,
\item $\forall y \in V^i$, $\pi^{-1}_+ (\pi^{-1}_S(y)) = \chi^{i*}_{Sy} P$.
\end{itemize}
$P(M,G) \to M$ is called the structure bundle of the principal composite bundle.
\end{defi}
(A good open cover of $R$ is a set of simply connected contractible open sets covering $R$). $\chi^{i*}_{Sy}$ is the map induced in the bundles by $\chi^{i}_{Sy} = \chi^{i}_{S|\pi^{-1}_S(y)}$ (the fibre diffeomorphism of $S(R,M)$ over $y$), i.e. it is the bundle isomorphism such that the following diagram commutes:
$$\begin{CD}
P_+ @<{\hookleftarrow}<< \pi^{-1}_+(\pi^{-1}_S(y)) @<{\chi^{i*}_{Sy}}<< P \\
@V{\pi_+}VV @V{\pi_+}VV @VV{\pi_P}V \\
S @<{\hookleftarrow}<< \pi^{-1}_S(y) @>{\chi^{i}_{Sy}}>> M \\
@V{\pi_S}VV @V{\pi_S}VV \\
R @<{\hookleftarrow}<< \{y\} 
\end{CD} $$

As a principal bundle defines a total space which is locally a cartesian product of manifolds, a principal composite bundle defines a total space which is locally a fibre bundle.

\begin{defi}[Total twisted bundle of a principal composite bundle]
Let $P_+(S,G) \xrightarrow{\pi_+} S(R,M) \xrightarrow{\pi_S} R$ be a principal composite $P(M,G)$-bundle. Let $\phi^\alpha_P : U^\alpha \times G \xrightarrow{\simeq} \pi_P^{-1}(U^\alpha) \subset P$ be the local trivialisation of $P(M,G)$ ($\{U^\alpha\}_\alpha$ being a good open cover of $M$). We call total twisted bundle of $P_+ \to S \to R$, the set of principal $G$-bundles $\{P_+^i(M\times V^i,G) \xrightarrow{\pi_{++}^i} M \times V^i\}_i$ defined by the following local trivialisation:
$$ \phi^{\alpha i}_{++} : \begin{array}{rcl} U^\alpha \times V^i \times G & \to & P_+^i \\ (x,y,g) & \mapsto & \chi^{i*}_{Sy} \phi^\alpha_P(x,g) \end{array} $$
where $P_+^i = \pi_+^{-1}(\pi_S^{-1}(V^i))$ with the projection defined by $\forall p \in P_+^i$, $\pi_{++}^i(p) = (\chi^{i}_{S\pi_S\circ \pi_+(p)} \circ \pi_+(p);\pi_S \circ \pi_+(p))$.
\end{defi}

Since $\chi^{i*}_{Sy} \phi^\alpha_P(s,G) = \pi_+^{-1}(\chi^{i-1}_{Sy}(x))$, for each $y \in V^i \cap V^j$, $\chi^{j*}_{Sy} \phi^\alpha_P(\chi^j_{Sy} \circ \chi^{i-1}_{Sy}(x),G) = \chi^{i*}_{Sy} \phi^\alpha_P(x,G)$.

\begin{defi}[Torsion functions of the twisted total bundle]
We call torsion functions of the total twisted bundle of a principal composite $P(M,G)$-bundle $P_+(S,G) \xrightarrow{\pi_+} S(R,M) \xrightarrow{\pi_S} R$, the automorphisms of $M$ defined for all $y \in V^i \cap V^j$ by
$$ \varphi^{ij}_y : \begin{array}{rcl} M & \to & M \\ x & \mapsto & \chi^j_{Sy} \circ \chi^{i-1}_{Sy}(x) \end{array} $$
$\varphi^{ij}_y$ represents the torsion of $\{P^i_+(M \times V^i,G)\}_i$ since
$$ \phi^{\alpha i}_{++}(x,y,G) = \phi^{\alpha j}_{++}(\varphi^{ij}_y(x),y,G) $$
\end{defi}
The total twisted bundle is a principal $G$-bundle if and only if $\varphi^{ij}_y = id_M$, i.e. if $S = R \times M$ (the bundle $S(R,M)$ is trivial).\\

\begin{figure}[h]
\begin{center}
\includegraphics[width=8cm]{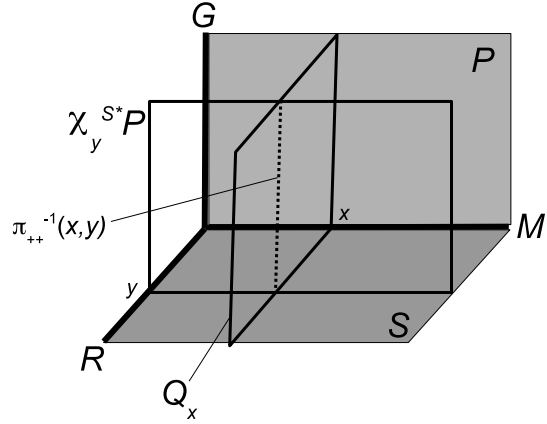}
\end{center}
\caption{\label{scheme} Scheme of a principal composite $P(M,G)$-bundle $P_+(S,G) \to S(R,M) \to R$. A single fibre over $(x,y)$ is represented by a dotted line. This fibre belongs to the longitudinal leaf of fibres $\chi_{Sy}^* P$ and to the transversal leaf of fibres $Q_x$.}
\end{figure}

There are three notions of fibre in a principal composite bundle (see fig. \ref{scheme}). One is the usual fibre of the total bundle of the composite, i.e. $\pi_{++}^{i-1}(x,y)$ for $(x,y) \in M \times V^i$. The second kind plays the role in the principal composite bundle of the fibre diffeomorphic to the structure group in a principal bundle, i.e. $\pi^{-1}_+(\pi^{-1}_S(y)) = \chi^{i*}_{Sy} P$. We call it a longitudinal leaf of fibres. The third kind has no analogue in a simple principal bundle; it is a transversal leaf of fibres:
\begin{defi}[Transversal bundles]
Let $P_+(S,G) \xrightarrow{\pi_+} S(R,M) \xrightarrow{\pi_S} R$ be a principal composite $P(M,G)$-bundle, and $\phi^{\alpha i}_{++}$ be the local trivialisation of its total twisted bundle. We call transerval bundle over $x \in U^\alpha \subset M$ the set of principal $G$-bundles $\{Q_x^{\alpha i}(V^i,G) \xrightarrow{\pi_{Qx}^\alpha} V^i \}_i $ defined by the following local trivialisations:
$$ \phi_{Qx}^{i(\alpha)} : \begin{array}{rcl} V^i \times G & \to & Q^{\alpha i}_x \subset P_+ \\ (y,g) & \mapsto & \phi^{\alpha i}_{++}(x,y,g) \end{array} $$
The projection of a transversal bundle is then defined by $\forall q \in Q_x^{\alpha i}$, $\pi_{Qx}^\alpha(q) = \pi_S \circ \pi_+(q)$.
\end{defi}

We remark that if $x \in U^\alpha \cap U^\beta$ then there exists two diffeomorphic transversal bundles over $x$ : $Q_x^{\alpha i} \simeq Q_x^{\beta i}$. We denote by $\phi_{Qx}^{\alpha \beta (i)} : Q_x^{\beta i} \xrightarrow{\simeq} Q_x^{\alpha i}$ this diffeomorphism. $\forall x \in U^\alpha \cap U^\beta \cap U^\gamma$ and for $q \in Q_x^{\gamma i}$, $\phi_{Q_x}^{\alpha \gamma (i)}(q) = \phi_{Q_x}^{\alpha \beta (i)} \circ \phi_{Q_y}^{\beta \gamma (i)}(q)\cdot h^{\alpha \beta \gamma (i)}_x(\pi_{Qx}^\alpha(q))$ with $h^{\alpha \beta \gamma(i)}_x(y) \in G$ (the dot in the preceeding expression denotes the right action of $G$ on $Q_x^{\gamma i}$). 
\begin{prop}
The family of the transerval bundles $\{Q_x^{\alpha i}(R,G)\to R \}_{x \in U^\alpha, \alpha}$ of a principal composite bundle is torsion free, i.e. $h^{\alpha \beta \gamma(i)}_x(y) = e$ ($e$ is the identity element of $G$).
\end{prop} 
This property follows from the following commutative diagram
$$ \begin{CD}
\pi_{Qx}^{\beta -1}(y) @>{\phi_{Qx}^{\alpha \beta(i)}}>> \pi_{Qx}^{\alpha -1}(y) \\
@V{\chi^{i*-1}_{Sy}}V{\simeq}V @V{\chi^{i*-1}_{Sy}}V{\simeq}V \\
\pi_P^{-1}(x) @>>{\phi_{Px}^{\alpha-1} \circ \phi_{Px}^{\beta}}> \pi_P^{-1}(x) 
\end{CD} $$
$\forall y \in V^i$.\\

Remark : Let $\{W^{\alpha i} \}_{\alpha,i}$ be the good open cover of $S$ defined by
$$ W^{\alpha i} = \bigcup_{y \in V^i} \chi^{i-1}_{Sy}(U^\alpha) $$
The local trivialisation of the principal $G$-bundle $P_+(S,G) \xrightarrow{\pi_+} S$ is
$$ \phi^{\alpha i}_+ : \begin{array}{rcl} W^{\alpha i} & \to & \pi^{-1}_+ (W^{\alpha i}) \\ (s,g) & \mapsto & \chi^{i*}_{S\pi_S(s)} \phi^\alpha_P(\chi^i_{S\pi_S(s)}(s),g) \end{array} $$

\subsection{Local data defining a principal composite bundle}
A principal bundle, as $P(M,G) \to M$, is totally determined by the knowledge of the manifolds $P$ and $M$ and of the local diffeomorphism $\phi^\alpha_P : U^\alpha \times G \to P$ (or equivalently the fibre diffeormophism $\phi^\alpha_{Px} = \phi^\alpha_P(x,.)$). However, it is often more practical to determine the structure of a principal bundle by using entities defined only with $M$ and $G$ (without any explicit reference to $P$). Indeed the explicit geometry of the manifold $P$ is often unknown. The local data defining a principal bundle are the transition functions $g_P^{\alpha \beta} \in \Omega^0(U^\alpha \cap U^\beta,G)$ (see ref. \cite{naka}) which are related to the local diffeomorphisms by
\begin{equation}
\forall x \in U^\alpha \cap U^\beta, \quad \phi_P^\beta(x,e) = \phi_P^\alpha(x,g^{\alpha \beta}_P(x))
\end{equation}
They satisfy the cocycle relations
\begin{eqnarray}
\forall x \in U^\alpha \cap U^\beta \cap U^\gamma, & \quad & g^{\alpha \beta}_P(x) g^{\beta \gamma}_P(x) = g^{\alpha \gamma}_P(x) \\
\forall x \in U^\alpha \cap U^\beta, & \qquad & g^{\alpha \beta}_P(x) = g^{\beta \alpha}_P(x)^{-1}
\end{eqnarray}

Since $\chi^{j*}_{Sy} \phi^\alpha_P(\varphi^{ij}_y(x),G) = \chi^{i*}_{Sy} \phi^\alpha_P(x,G)$, there exists $g^{ij(\alpha)}_{Qx} \in \Omega^0(V^i \cap V^j,G)$ such that
$$ \chi^{j*}_{Sy} \phi^\alpha_P(\varphi^{ij}_y(x),e) = \chi^{i*}_{Sy} \phi^\alpha_P(x,g^{ij(\alpha)}_{Qx}(y)) $$

\begin{prop}
$\forall x \in U^\alpha \cap U^\beta$ and $\forall y \in V^i \cap V^j$, we have
\begin{equation}
g^{\alpha \beta}_P(x) g^{ij(\beta)}_{Qx}(y) = g^{ij(\alpha)}_{Qx}(y) g^{\alpha \beta}_P(\varphi^{ij}_y(x))
\end{equation}
\end{prop}
This property follows from the two following calculations:
\begin{eqnarray}
\chi^{j*}_{Sy} \phi^\beta_P(\varphi^{ij}_y(x),e) & = & \chi^{j*}_{Sy} \phi^\alpha_P(\varphi^{ij}_y(x),g^{\alpha \beta}_P(\varphi^{ij}_y(x))) \\
& = & \chi^{i*}_{Sy} \phi^\alpha_P(x,g^{ij(\alpha)}_{Qx}(y) g^{\alpha \beta}_P(\varphi^{ij}_y(x)))
\end{eqnarray}
\begin{eqnarray}
\chi^{j*}_{Sy} \phi^\beta_P(\varphi^{ij}_y(x),e) & = & \chi^{i*}_{Sy} \phi^\beta_P(x,g^{ij(\beta)}_{Qx}(y)) \\
& = & \chi^{i*}_{Sy} \phi^\alpha_P(x,g^{\alpha \beta}_P(x)g^{ij(\beta)}_{Qx}(y))
\end{eqnarray}

 The transition functions of the total twisted  bundle of a principal composite $P(M,G)$-bundle $P_+(S,G) \to S(R,M) \to R$ are defined by
\begin{equation}
\forall (x,y) \in U^\alpha\cap U^\beta \times V^i \cap V^j, \quad g^{(\alpha i)(\beta j)}_{++}(x,y) = g^{\alpha \beta}_P(x)g^{ij(\beta)}_{Qx}(y) = g^{ij(\alpha)}_{Qx}(y) g^{\alpha \beta}_P(\varphi^{ij}_y(x))
\end{equation}
They are related to the local diffeomorphisms by
\begin{equation}
\forall (x,y) \in U^\alpha \cap U^\beta \times V^i \cap V^j, \quad \phi^{\beta j}_{++}(\varphi^{ij}_y(x),y,e) = \phi^{\alpha i}_{++}(x,y,g^{(\alpha i)(\beta j)}_{++}(x,y))
\end{equation}

\begin{prop}
The transversal transition functions satisfy the twisted cocycle relation
\begin{equation}
\forall y \in V^i \cap V^j \cap V^k, \quad g^{ik(\alpha)}_{Qx}(y) = g^{ij(\alpha)}_{Qx}(y) g^{jk(\alpha)}_{Q \varphi^{ij}_y(x)}(y)
\end{equation}
\end{prop}
This property follows from the two following calculations:
\begin{equation}
\chi^{k*}_{Sy} \phi^\alpha_P(\varphi^{ik}_y(x),e) = \chi^{i*}_{Sy} \phi^\alpha_P(x,g^{ik(\alpha)}_{Qx}(y))
\end{equation}
\begin{eqnarray}
\chi^{k*}_{Sy} \phi^\alpha_P(\varphi^{ik}_y(x),e) & = & \chi^{k*}_{Sy} \phi^\alpha_P(\varphi^{jk}_y \circ \varphi^{ij}_y(x),e) \\
& = & \chi^{j*}_{Sy} \phi^\alpha_P(\varphi^{ij}_{Sy}(x),g^{jk(\alpha)}_{Q\varphi^{ij}_y(x)}(y)) \\
& = & \chi^{i*}_{Sy} \phi^\alpha_P(x,g^{ij(\alpha)}_{Qx}(y) g^{jk(\alpha)}_{Q\varphi^{ij}_y(x)}(y))
\end{eqnarray}

\begin{prop}
The transition functions of the total twisted bundle satisfy the twisted cocycle relation
\begin{equation}
\forall (x,y) \in U^\alpha \cap U^\beta \cap U^\gamma \times V^i \cap V^j \cap V^k, \quad g^{(\alpha i)(\gamma k)}_{++}(x,y) = g^{(\alpha i)(\beta j)}_{++}(x,y) g^{(\beta j)(\gamma k)}_{++}(\varphi^{ij}_y(x),y)
\end{equation}
\end{prop}
Indeed we have
\begin{eqnarray}
g^{(\alpha i)(\gamma k)}_{++}(x,y) & = & g^{\alpha \gamma}_P(x) g^{ik(\gamma)}_{Qx}(y) \\
& = & g^{\alpha \beta}_P(x) g^{\beta \gamma}_P(x) g^{ij(\gamma)}_{Qx}(y) g^{jk(\gamma)}_{Q\varphi^{ij}_y(x)}(y) \\
& = & g^{\alpha \beta}_P(x) g^{ij(\beta)}_{Qx}(y) g^{\beta \gamma}_P(\varphi^{ij}_y(x)) g^{jk(\gamma)}_{Q\varphi^{ij}_y(x)}(y)  \\
& = & g^{(\alpha i)(\beta j)}_{++}(x,y) g^{(\beta j)(\gamma k)}_{++}(\varphi^{ij}_y(x),y)
\end{eqnarray}

The twisted cocycle relation can be rewritten as
\begin{equation}
g^{(\alpha i)(\gamma k)}_{++}(x,y) = g^{(\alpha i)(\beta j)}_{++}(x,y) g^{(\beta j)(\gamma k)}_{++}(x,y) h^{(\alpha i)(\beta j)(\gamma k)}(x,y)
\end{equation}
with
\begin{equation}
h^{(\alpha i)(\beta j)(\gamma k)}(x,y) = g^{(\beta j)(\gamma k)}_{++}(x,y)^{-1} g^{(\beta j)(\gamma k)}_{++}(\varphi^{ij}_y(x),y)
\end{equation}
$h^{(\alpha i)(\beta j)(\gamma k)} \in \Omega^0(U^\alpha \cap U^\beta \cap U^\gamma \times V^i \cap V^j \cap V^k,G)$ represents the obstruction to lift the total twisted bundle into a principal bundle. The notion of total twisted bundle introduced in this paper extends the notions of twisted bundle and bundle gerbes ref. \cite{mackaay, aschieri, kalkkinen, baez}. 

\begin{propo}
Let $g^{(\alpha i)(\beta j)}_+ \in \Omega^0(W^{\alpha i}\cap W^{\beta j},G)$ be the transition functions of the principal $G$-bundle $P_+(S,G)$. These transition functions are related to the transition function of the total twisted bundle by
\begin{equation}
g^{(\alpha i)(\beta j)}_{++}(x,y) = g^{(\alpha i)(\beta j)}_+(\chi^{i-1}_{Sy}(x))
\end{equation}
\end{propo}
Indeed $\forall s \in W^{\alpha i}\cap W^{\beta j}$ we have
\begin{equation}
\phi^{\beta j}_+(s,e) = \phi^{\alpha i}_+(s,g^{(\alpha i)(\beta j)}_+(s))
\end{equation}
and
\begin{eqnarray}
\phi^{\beta j}_{++}(\varphi^{ij}_y(x),y,e) & = & \phi^{\beta j}_+(\chi^{i-1}_{Sy}(x),e) \\
\phi^{\alpha i}_{++}(x,y,g) & = & \phi^{\alpha i}_+(\chi^{i-1}_{Sy}(x),g)
\end{eqnarray}
We can note that the usual cocycle relation concerning $g^{(\alpha i)(\beta j)}_+$ involves the twisted cocycle relation concerning $g^{(\alpha i)(\beta j)}_{++}$ :
\begin{equation}
g^{(\alpha i)(\beta j)}_+(\chi^{i-1}_{Sy}(x)) g^{(\beta j)(\gamma k)}_+( \underbrace{\chi^{i-1}_{Sy}(x)}_{\chi^{j-1}_{Sy} \circ \varphi^{ij}_y(x)}) = g^{(\alpha i)(\gamma k)}_+(\chi^{i-1}_{Sy}(x))
\end{equation}

\begin{propo}
A composite principal $P(M,G)$-bundle over $S(M,R) \to R$ is completely determined by the knowledge of automorphisms $\varphi^{ij}_x : M \xrightarrow{\simeq} M$ and functions $g^{(\alpha i)(\beta j)}_{++} \in \Omega^0(U^\alpha \cap U^\beta \times V^i \cap V^j,G)$, such that $g^{(\alpha i)(\beta i)}_{++}(x,y)$ is independent of $y \in R$ and of $i$, and satisfying the twisted cocycle relations
\begin{equation}
\forall (x,y) \in U^\alpha \cap U^\beta \cap U^\gamma \times V^i\cap V^j \cap V^k,  \quad  g^{(\alpha i)(\beta j)}_{++}(x,y) g^{(\beta j)(\gamma k)}_{++}(\varphi^{ij}_y(x),y) = g^{(\alpha i)(\gamma k)}_{++}(x,y)
\end{equation}
\end{propo}
Indeed, we can reconstruct the manifold $S$ by
\begin{equation}
S = \bigsqcup_i V^i \times M/\sim \qquad (y,x) \sim (y',x') \iff y=y' \text{ and } y \in V^i \cap V^j \Rightarrow x'=\varphi^{ij}(x) 
\end{equation}
and the manifold $P_+$ by
\begin{equation}
P_+ = \bigsqcup_{\alpha,i} W^{\alpha i} \times G / \sim \qquad (s,g) \sim (s',g') \iff s=s' \text{ and } s \in W^{\alpha i}\cap W^{\beta j} \Rightarrow g'=g g_{++}^{(\alpha i)(\beta j)}(\chi^i_{S \pi_S(s)}(s),\pi_S(s))
\end{equation}

\section{Composite connections and composite holonomies}
\subsection{Global definition of a composite connection}
The fibers of $P_+$ being diffeomorphic to $G$, there exists a tangent vector subspace such that $T_rP_+ \supset V_r P_+ \simeq \mathfrak g$ for $r \in P_+$ (where $\mathfrak g$ is Lie algebra of $G$) called the space of vertical tangent vectors (see ref. \cite{naka}). A connection is a choice of a supplementary subspace $H_rP_+$ called the space of horizontal tangent vectors, $T_rP_+ = V_rP_+ \oplus H_rP_+$. We want to reduce the choice of the horizontal tangent space to be compatible with the composite structure.
\begin{defi}[Composite connection]
Let $P_+(S,G) \to S(R,M) \to M$ be a principal composite $P(M,G)$-bundle. A composite connection is a choice of a horizontal tangent space $H_rP_+$ at each point $r$ of $P_+$ such that there exists a horizontal tangent space $H_p P$ at each point $p$ of $P$ with $\chi^{i*}_{Sy*} H_p P \subset H_{\chi^{i*}_{Sy}(p)}P_+$.
\end{defi}
$\chi^{i*}_{Sy*}$ denotes the tangent map of $\chi^{i*}_{Sy}$ (the lower star denotes the push-forward).\\
A connection being defined by a 1-form (for which the horizontal tangent space is its kernel, see ref. \cite{naka}), we have the following definition.
\begin{defi}[Composite connection 1-form]
Let $P_+(S,G) \to S(R,M) \to M$ be a principal composite $P(M,G)$-bundle. A connection 1-form $\omega \in \Omega^1(P_+,\mathfrak g)$ is a composite principal connection 1-form if there exists a connection 1-form $\omega_P \in \Omega^1(P,\mathfrak g)$ such that $\forall y \in V^i$, $\chi^{i**}_{Sy} i^*_y \omega = \omega_P$ where $i_y : \pi_+^{-1}(\pi_S^{-1}(y)) \xrightarrow{\hookrightarrow} P_+$ is the canonical injection.
\end{defi}
$\chi^{i**}_{Sy}$ denotes the cotangent map of $\chi^{i*}_{Sy}$ (the second upper star denotes the pull-back). The important property of a composite connection 1-form is that $\chi^{i**}_{Sy} i^*_y \omega$ is independent of $y$ and of $i$.\\
Let $i^\alpha_x : Q^\alpha_x \xrightarrow{\hookrightarrow} P_+$ be the canonical injection. The transversal bundle is endowed with the connection 1-form $\omega^\alpha_{Qx} = i^{\alpha *}_x \omega \in \Omega^1(Q^\alpha_x, \mathfrak g)$.

\subsection{Local data associated with a composite connection}
Let $\omega \in \Omega^1(P_+,\mathfrak g)$ be a composite connection of $P_+(S,G) \to S(R,M) \to R$. Let $\sigma^{\alpha i}_{M \times R} \in \Gamma(U^\alpha \times V^i,P_+)$ be a local section of the principal bundle $P_+^i(M\times V^i,G)$, such that 
\begin{equation}
\forall (x,y) \in U^\alpha \cap U^\beta \times V^i \cap V^j,\quad \sigma^{\beta j}_{M \times R}(\varphi^{ij}_y(x),y) = \sigma^{\alpha i}_{M \times R}(x,y) \cdot g^{(\alpha i)(\beta j)}_{++}(x,y)
\end{equation}
The gauge potential associated with the connection is $A^{\alpha i}_{+} = \sigma^{\alpha i *}_{M \times R} \omega \in \Omega^1(U^\alpha \times V^i, \mathfrak g)$. By construction we have 
\begin{eqnarray}
& & \forall (x,y) \in U^\alpha \cap U^\beta \times V^i \cap V^j, \nonumber \\
& & \quad \varphi^{ij*} A^{\beta j}_{+}(x,y) = g^{(\alpha i)(\beta j)}_{++}(x,y)^{-1} A^{\alpha i}_{+}(x,y) g^{(\alpha i)(\beta j)}_{++}(x,y) + g^{(\alpha i)(\beta j)}_{++}(x,y)^{-1} d_{M \times R} g^{(\alpha i)(\beta j)}_{++}(x,y)
\end{eqnarray}

where $\varphi^{ij*}: \Omega^*(M \times V^i \cap V^j) \to \Omega^*(M \times V^i \cap V^j)$ denotes the cotangent map of $(x,y) \mapsto \varphi^{ij}_y(x)$. We have then
\begin{equation}
\varphi^{ij*} A^{\beta j}_+(x,y) = A^{\beta j}_{+ \nu}(\varphi^{ij}_y(x),y) \frac{\partial \varphi^{ij \nu}_y(x)}{\partial x^\mu} dx^\mu + \left(A^{\beta j}_{+\nu}(\varphi^{ij}_y(x),y) \frac{\partial \varphi^{ij \nu}_y(x)}{\partial y^a} + A^{\beta j}_{+a}(\varphi^{ij}_y(x),y) \right)dy^a
\end{equation}
where $\varphi^{ij \nu}_y(x)$ is the $\nu$-th coordinates of the point $\varphi^{ij}_y(x) \in M$.\\

Let $F^{\alpha i}_+ = dA^{\alpha i}_+ + A^{\alpha i}_+ \wedge A^{\alpha i}_+ \in \Omega^2(U^\alpha \times V^i,\mathfrak g)$ be the curvature of the principal composite bundle. By construction we have
\begin{equation}
\varphi^{ij*} F^{\beta j}_+(x,y) = g^{(\alpha i)(\beta j)}_{++}(x,y)^{-1} F^{\alpha i}_+(x,y) g^{(\alpha i)(\beta j)}_{++}(x,y)
\end{equation}

We can also use the language of the 2-connections on the total twisted bundle (see ref. \cite{mackaay, aschieri, kalkkinen, baez}) :
\begin{equation}
A^{\beta j}_+(x,y) = g^{(\alpha i)(\beta j)}_{++}(x,y)^{-1} A^{\alpha i}_{+}(x,y) g^{(\alpha i)(\beta j)}_{++}(x,y) + g^{(\alpha i)(\beta j)}_{++}(x,y)^{-1} d_{M \times R} g^{(\alpha i)(\beta j)}_{++}(x,y) + A^{ij(\beta)}_{++}(x,y)
\end{equation}
where $A^{ij(\beta)}_{++}(x,y) = A^{\beta j}_+(x,y) - \varphi^{ij*} A^{\beta j}_+(x,y)$ is the potential of the 2-connection. By construction we have
\begin{eqnarray}
& & F^{\alpha j}_+(x,y) - g^{ij(\alpha)}_{Qx}(y)^{-1} F^{\alpha i}_+(x,y) g^{ij(\alpha)}_{Qx}(y) \nonumber \\
& & \qquad  =  F^{\alpha j}_+(x,y) - \varphi^{ij*} F^{\alpha j}_+(x,y) \\
& & \qquad  = d_{M \times R} A^{ij(\alpha)}_{++}(x) + [A^{\alpha j}_+(x,y), A^{ij(\alpha)}_{++}(x,y)] - A^{ij(\alpha)}_{++}(x,y) \wedge A^{ij(\alpha)}_{++}(x,y)
\end{eqnarray}
$F^{\alpha i}_+$ plays then the role of the curving associated with the 2-connection (usually denoted by $B$).

\subsection{The intertwining curvature}
The transversal bundle $Q_x^{\alpha i}$ is endowed with the gauge potential $A^{i(\alpha)}_{Qx} = \sigma^{\alpha i}_{M \times R}(x,.)^* \omega^\alpha_{Qx} = j^*_x A^{\alpha i}_+ \in \Omega^1(V^i,\mathfrak g)$ where $j_x: \begin{array}{rcl} R & \to & M \times R \\ y & \mapsto & (x,y) \end{array}$. We have then
\begin{equation}
 \forall y \in V^i \cap V^j, \quad \varphi^{ij*} A^{j(\alpha)}_{Qx}(y) = g^{ij(\alpha)}_{Qx}(y)^{-1} A^{i(\alpha)}_{Qx}(y) g^{ij(\alpha)}_{Qx}(y) + g^{ij(\alpha)}_{Qx}(y)^{-1} d_R g^{ij(\alpha)}_{Qx}(y)
\end{equation}
For $x \in U^\alpha \cap U^\beta$, the gauge potentials of $Q_x^{\alpha i}(R,G)$ and $Q^{\beta i}_x(R,G)$ are related by
\begin{equation}
A^{i(\beta)}_{Qx}(y) = g^{\alpha \beta}_P(x)^{-1} A^{i(\alpha)}_{Qx}(y) g^{\alpha \beta}_P(x)
\end{equation}
$\chi^{i*-1}_{Sy} \sigma^{\alpha i}_{M \times R}(.,y) \in \Gamma(U^\alpha,P)$ is a $(y,i)$-dependent local section of the principal bundle $P(M,G)$. With it we can define a $(y,i)$-dependent gauge potential of $P$, $A^{\alpha(i)}_{Py} = \sigma^{\alpha i *}_{M \times R} \chi^{i*-1 *}_{Sy} \omega_P = \sigma^{\alpha i *}_{M \times R} \chi^{i*-1 *}_{Sy} \chi^{i**}_{Sy} i^*_y \omega = j^*_y A^{\alpha i}_+ \in \Omega^1(U^\alpha,\mathfrak g)$ where $j_y : \begin{array}{rcl} M & \to & M \times R \\ x & \mapsto & (x,y) \end{array}$. It is more interesting to endow $P(M,G)$ with a gauge potential independent of $y$. Let $\sigma_M^\alpha \in \Gamma(U^\alpha,P)$ be a local section of $P$ such that $\sigma^\beta_M(x) = \sigma^\alpha_M(x) \cdot g^{\alpha \beta}_P(x)$, we define the gauge potential $A^\alpha_P = \sigma^{\alpha *}_M \omega_P \in \Omega^1(U^\alpha,\mathfrak g)$. The relation between $A^\alpha_P$ and the family $\{A^{\alpha (i)}_{Py} \}_{y \in V^i,i}$ is just a $(y,i)$-dependent gauge transformation, indeed
\begin{equation}
\forall (x,y) \in U^\alpha \times V^i, \exists g^{\alpha i}_y(x) \in G, \quad \chi^{i*-1}_{Sy} \sigma^{\alpha i}_{M \times R}(x,y) = \sigma^\alpha_M(x) \cdot g^{\alpha i}_y(x)
\end{equation}
and then
\begin{equation}
A^{\alpha (i)}_{Py}(x) = g^{\alpha i}_y(x)^{-1} A^\alpha_P(x) g^{\alpha i}_y(x) + g^{\alpha i}_y(x) d_M g^{\alpha i}_{y}(x)
\end{equation}
We note that
\begin{equation}
\forall (x,y) \in U^\alpha\cap U^\beta \times V^i, \quad g^{\alpha \beta}_P(x) g^{\beta i}_y(x) = g^{\alpha i}_y(x) g^{\alpha \beta}_P(x)
\end{equation}
We have then
\begin{eqnarray}
\forall x \in U^\alpha \cap U^\beta, & \quad & A^\beta_P(x) = g^{\alpha \beta}_P(x)^{-1} A^\alpha_P(x) g^{\alpha \beta}_P(x) + g^{\alpha \beta}_P(x)^{-1} d_M g^{\alpha \beta}_P(x) \\
& & A^{\beta(i)}_{Py}(x) = g^{\alpha \beta}_P(x)^{-1} A^{\alpha(i)}_{Py}(x) g^{\alpha \beta}_P(x) + g^{\alpha \beta}_P(x)^{-1} d_M g^{\alpha \beta}_P(x)
\end{eqnarray}
The gauge potentials of the total, structure and transversal bundles are related by
\begin{eqnarray}
A^{\alpha i}_+(x,y) & = & A^{\alpha (i)}_{Py}(x) + A^{i(\alpha)}_{Qx}(y) \\
& = & g^{\alpha i}_y(x)^{-1} A^\alpha_P(x) g^{\alpha i}_y(x) + A^{i(\alpha)}_{Qx}(y) + g^{\alpha i}_y(x)^{-1} d_M g^{\alpha i}_y(x)
\end{eqnarray}
We prefer another gauge choice : $\tilde A^{\alpha i}_+(x,y) = g^{\alpha i}_y(x) A^{\alpha i}_+(x,y) g^{\alpha i}_y(x)^{-1} + g^{\alpha i}_y(x) d_{M \times R} g^{\alpha i}_y(x)^{-1}$, in which we have
\begin{equation}
\tilde A^{\alpha i}_+(x,y) = A^\alpha_P(x) + \tilde A^{i(\alpha)}_{Qx}(y)
\end{equation}
with $\tilde A^{i(\alpha)}_{Qx}(y) = g^{\alpha i}_y(x) A^{i(\alpha)}_{Qx}(y) g^{\alpha i}_y(x)^{-1} + g^{\alpha i}_y(x) d_R g^{\alpha i}_y(x)^{-1}$. We call this choice the gauge of decomposition since it isolates the gauge potential of $P(M,G)$.\\
Finally, we compute the curvature of the composite bundle.
\begin{eqnarray}
\tilde F^{\alpha i}_+ & = & d_{M \times R} \tilde A^{\alpha i}_+ + \tilde A^{\alpha i}_+ \wedge \tilde A^{\alpha i}_+ \\
& = & F^\alpha_P + \tilde F^{i (\alpha)}_{Qx} + \underbrace{d_M \tilde A^{i(\alpha)}_{Qx} + \left[A^\alpha_P , \tilde A^{i(\alpha)}_{Qx} \right]}_{D^\alpha_P \tilde A^{i(\alpha)}_{Qx}}
\end{eqnarray}
where $D^\alpha_P = d_M +[A^\alpha_P,.]$ is the covariant differential associated with the connection of $P(M,G)$, $F^\alpha_P$ is the curvature of $P(M,G)$ and $\tilde F^{i(\alpha)}_{Qx}$ is the curvature of $Q^{\alpha i}_x(V^i,G)$. We note that
\begin{equation}
\forall (x,y)\in U^\alpha \cap U^\beta \times V^i \cap V^j, \quad \varphi^{ij*} \tilde F^{\beta j}_+ = \tilde g^{(\alpha i)(\beta j)-1}_{++} \tilde F^{\alpha i}_+ \tilde g^{(\alpha i)(\beta j)}_{++} \text { and } \varphi^{ij*} D^\beta_P \tilde A^{j(\beta)}_{Qx}= \tilde g^{(\alpha i)(\beta j)-1}_{++} D^\alpha_P \tilde A^{i(\alpha)}_{Qx} \tilde g^{(\alpha i)(\beta j)}_{++}
\end{equation}
with $\tilde g^{(\alpha i)(\beta j)}_{++}(x,y) = g^{\alpha i-1}_y(\varphi^{ij}_y(x)) g^{(\alpha i)(\beta j)}_{++}(x,y) g^{\beta j}_y(x)$. We call $D^\alpha_P \tilde A^{i(\alpha)}_{Qx}$ the intertwining curvature of the composite bundle, since it measures the covariant variations of the connection of $Q_x^{\alpha i}(V^i,G)$ with respect to the variations of $x$ (the covariance being defined with respect to the connection of $P(M,G)$).  

\subsection{Composite holonomy}
Let $\mathcal C_{M \times R} \in \mathcal L_{(x_0,y_0)}(M \times V^i)$ be a closed path in $M \times V^i$, $\mathcal C_M \in \mathcal L_{x_0}M$ be its ``image'' in $M$, and $\mathcal C_R \in \mathcal L_{y_0}V^i$ be its ``image'' in $R$. The holonomy of the path $\mathcal C_{M \times R}$ in the total bundle $P_+^i(M\times V^i,G)$ measures the difference along the fibre $\pi_{++}^{i-1}(x_0,y_0)$ between the two endpoints of the open horizontal lift of the closed path $\mathcal C_{M \times R}$. It is defined by (see ref. \cite{naka})
\begin{equation}
\hol_{\tilde A_+}(\mathcal C_{M \times R}) = \P_{\mathcal C_{M \times R}}e^{\oint \tilde A_+}
\end{equation}
where $\P_{\mathcal C_{M \times R}}$ is the path-ordering operator along $\mathcal C_{M \times R}$, i.e. $\P_{\mathcal C}e^{\int_{x_0}^{x} A}$ is the solution of the following equation
\begin{equation}
\frac{d \P_{\mathcal C}e^{\int_{x_0}^{x(s)} A}}{ds} = \P_{\mathcal C}e^{\int_{x_0}^{x(s)} A} A_\mu \frac{dx^\mu(s)}{ds}
\end{equation}
where $s \mapsto x(s)$ is a parametrization of $\mathcal C$.\\
In the composite bundle, we are interested in the comparison between the holonomy in the total bundle $\hol_{\tilde A_+}(\mathcal C_{M \times R})$ and the holonomies in the structure bundle $\hol_{A_P}(\mathcal C_M)$ and in a relevant transversal bundle (for example the transversal bundle at the base point $x_0$) $\hol_{\tilde A_{Qx_0}}(\mathcal C_R)$. Ideally, we would prefer the composite holonomy to be the composition of the component holonomies : $\hol_{\tilde A_+}(\mathcal C_{M \times R}) = \hol_{\tilde A_{Qx_0}}(\mathcal C_R) \hol_{A_P}(\mathcal C_M)$ (the product between the two holonomies is the group law of $G$). Obviously this is NOT the case. The following theorem expresses the difference between the composite holonomy and the composition of the component holonomies by using the intertwining curvature (which measures precisely the intertwining between the two connections).

\begin{theo}
Let $P_+(S,G) \to S(R,M) \to R$ be a principal composite $P(M,G)$-bundle endowed with a composite connection defined by the gauge potential $\tilde A_+^{\alpha i}(x,y) = A^\alpha_P(x) + \tilde A_{Qx}^{i(\alpha)}(y)$ (within the gauge of decomposition). Let $\mathcal C_R \in \mathcal L_{y_0} R$. Let $h\in \Gamma(\mathcal C_R,S)$ be a local section of $S(R,M)$ over $\mathcal C_R$. We suppose that
\begin{itemize}
\item there exists a local chart $V^i \subset R$ such that $\mathcal C_R \subset V^i$,
\item there exists a local chart $U^\alpha \subset M$ such that $\chi^i_Sh(\mathcal C_R) \subset U^\alpha$,
\end{itemize}
The difference between the composite holonomy of the section $h$ in the principal composite bundle (which is defined as being the holonomy of $\mathcal C_{M \times R}$ in the total bundle) and the composition of the holonomies of $\mathcal C_R$ and of $\mathcal C_M$ is
\begin{equation}
\label{theorem}
\hol_{\tilde A_+^{\alpha i}}(\mathcal C_{M \times R}) \hol_{A^\alpha_P}(\mathcal C_M)^{-1} \hol_{\tilde A^{i(\alpha)}_{Qx_0}}(\mathcal C_R)^{-1} = \P_{\mathcal C_R}e^{\oint \int_{x_0}^{\chi_S^ih(y)} T(x,y) D^\alpha_{P \mu} \tilde A^{i(\alpha)}_{Qxa}(y) T(x,y)^{-1} dx^\mu dy^a}
\end{equation}
where $x_0 = \chi^i_S h(y_0)$, $\mathcal C_M = \chi^i_S h(\mathcal C_R) \in \mathcal L_{x_0}M$, and $\mathcal C_{M \times R} = \{(\chi^i_Sh(y),y); y \in \mathcal C_R \} \in \mathcal L_{(x_0,y_0)} (M \times R)$. The second integral is along $\mathcal C_M$ (it is not ordered) and
\begin{equation}
T(x,y) = \P_{\mathcal C_R}e^{\int_{y_0}^y \tilde A_{Qx_0}^{i(\alpha)}} \P_{\mathcal C_M} e^{\int_{x_0}^x A_P^\alpha}
\end{equation}
\end{theo}
This theorem can be viewed as an equivalent of the non-abelian Stokes theorem (ref. \cite{karp}) in the composite bundle, and it is proved in the appendix. The term appearing in the r.h.s. of the equation (\ref{theorem}) is gauge equivariant : if
\begin{equation}
\tilde A^{\alpha i \prime}_+(x,y) = g(x,y)^{-1} \tilde A^{\alpha i \prime}_{+}(x,y) g(x,y) + g(x,y)^{-1} d_{M \times R} g(x,y)
\end{equation}
then
\begin{equation}
\P_{\mathcal C_R}e^{\oint \int_{x_0}^{\chi_S^ih(y)} T'(x,y) D^{\alpha \prime}_{P \mu} \tilde A^{i(\alpha) \prime}_{Qxa} T'(x,y)^{-1} dx^\mu dy^a} = g(x_0,y_0)^{-1} \P_{\mathcal C_R}e^{\oint \int_{x_0}^{\chi_S^ih(y)} T(x,y) D^\alpha_{P \mu} \tilde A^{i(\alpha)}_{Qxa} T(x,y)^{-1} dx^\mu dy^a} g(x_0,y_0)
\end{equation}
The composite holonomy is the composition of the component holonomies if and only if the intertwining curvature vanishes $D^{\alpha}_P \tilde A^{i(\alpha)}_{Qx} = d_M \tilde A^{i(\alpha)}_{Qx} + [A_P^\alpha, \tilde A^{i(\alpha)}_{Qx}]=0$.\\

If we relax the second assumption by supposing that $\mathcal C_M$ crosses several charts $U^\alpha$, the works of Alvarez in ref. \cite{alvarez} show that the correct definition of the holonomy of $\mathcal C_{M \times R}$ is
\begin{eqnarray}
\hol_{\tilde A_+}(\mathcal C_{M \times R}) & = & \P_{\mathcal C_{M \times R}}e^{\int_{(x_0,y_0)}^{(x^{\alpha \beta},y^{\alpha \beta})} \tilde A^{\alpha i}_+} g^{\alpha \beta}_P(x^{\alpha \beta}) \P_{\mathcal C_{M \times R}}e^{\int_{(x^{\alpha \beta},y^{\alpha \beta})}^{(x^{\beta \gamma},y^{\beta \gamma})}\tilde A^{\beta i}_+} g^{\beta \gamma}_P(x^{\beta \gamma}) ... \nonumber \\
& & ... g^{\zeta \alpha}_P(x^{\zeta \alpha}) \P_{\mathcal C_{M \times R}}e^{\int_{(x^{\zeta \alpha},y^{\zeta \alpha})}^{(x_0,y_0)} \tilde A^{\alpha i}_+}
\end{eqnarray}
Where $U^\alpha,U^\beta,...,U^\zeta$ are the charts crossed by $\mathcal C_M$, $x^{\alpha \beta}$ is an arbitrary point on $U^\alpha \cap U^\beta \cap \mathcal C_M$ and $y^{\alpha \beta}$ is the point of $\mathcal C_R$ such that $\chi_S^i h(y^{\alpha \beta}) = x^{\alpha \beta}$. The preceeding expression is independant of the choice of the points $\{x^{\alpha \beta} \}_{\alpha,\beta}$. We can then apply theorem 1 on each piece of $\mathcal C_{M \times R}$, i.e.
\begin{eqnarray}
\hol_{\tilde A_+}(\mathcal C_{M \times R}) & = & \P_{\mathcal C_R}e^{\int_{y_0}^{y^{\alpha \beta}} \int_{x_0}^{\chi^i_Sh(y)} T^{\alpha i}(x,y) D^\alpha_P \tilde A^{i (\alpha)}_{Qx} T^{\alpha i}(x,y)^{-1}} \P_{\mathcal C_R}e^{\int_{y_0}^{y^{\alpha \beta}} \tilde A^{i(\alpha)}_{Qx_0}} \P_{\mathcal C_M}e^{\int_{x_0}^{x^{\alpha \beta}} A^\alpha_P} g^{\alpha \beta}_P(x^{\alpha \beta}) \nonumber \\
& & \times \P_{\mathcal C_R}e^{\int_{y^{\alpha \beta}}^{y^{\beta \gamma}} \int_{x^{\alpha \beta}}^{\chi^i_Sh(y)} T^{\beta i}(x,y) D^\beta_P \tilde A^{i(\beta)}_{Qx} T^{\beta i}(x,y)^{-1}} \P_{\mathcal C_R}e^{\int_{y^{\alpha \beta}}^{y^{\beta \gamma}} \tilde A^{i(\beta)}_{Qx^{\alpha \beta}}} \P_{\mathcal C_M}e^{\int_{x^{\alpha \beta}}^{x^{\beta \gamma}} A^\beta_P} g_P^{\beta \gamma}(x^{\beta \gamma}) \nonumber \\
& & ... \nonumber \\
& &  \times \P_{\mathcal C_R}e^{\int_{y^{\zeta \alpha}}^{y_0} \int_{x^{\zeta \alpha}}^{\chi_S^ih(y)} T^{\alpha i}(x,y) D^\alpha_P \tilde A^{i(\alpha)}_{Q x} T^{\alpha i}(x,y)^{-1}} \P_{\mathcal C_R}e^{\int_{y^{\zeta \alpha}}^{y_0} \tilde A^{i(\alpha)}_{Qx^{\zeta \alpha}}} \P_{\mathcal C_M}e^{\int_{x^{\zeta \alpha}}^{x_0} A^\alpha_P} \label{holocomp1}
\end{eqnarray}
The simple composition of the component holonomies does not appear because of the higher degree of intertwining due to the chart transitions.\\

Now we relax the first assumption by supposing that $\mathcal C_R$ crosses several charts $V^i$. Let $\mathcal C_M^i = \{ \chi^i_Sh(y), y \in \mathcal C_R \cap V^i \}$. $\mathcal C_M$ and $\mathcal C_{M \times R}$ are now two collections of disconnected paths. Let $y^{ij}$ be an arbitrary point on $V^i \cap V^j \cap \mathcal C_R$, $x^{ij(i)} = \chi^i_S h(y^{ij})$ and $x^{ij(j)} = \chi^j_Sh(y^{ij})$ be the images of $y^{ij}$ on the two disconnected paths $\mathcal C^i_M$ and $\mathcal C^j_M$. We note that $\varphi^{ij}(x^{ij(i)}) = x^{ij(j)}$. The direct generalization of the Alvarez formula:
\begin{equation}
\P_{\mathcal C_{M \times R}^i} e^{\int^{(x^{ij(i)},y^{ij})} \tilde A^{\alpha i}_+} \tilde g^{(\alpha i)(\alpha j)}_{++}(x^{ij(i)},y^{ij}) \P_{\mathcal C_{M \times R}^j}e^{\int_{(x^{ij(j)},y^{ij})} \tilde A^{\alpha j}_+}
\end{equation}
is well defined. Indeed it is independent of the arbitrary point $y^{ij}$. Let $\hat y^{ij}$ be another arbitrary point on $V^i \cap V^j \cap \mathcal C_R$. We have then
\begin{equation}
\P_{\mathcal C_{M \times R}^j}e^{\int_{(\hat x^{ij(j)},\hat y^{ij})} \tilde A^{\alpha j}_+} =  \P_{\mathcal C_{M \times R}^j}e^{\int_{(\hat x^{ij(j)},\hat y^{ij})}^{( x^{ij(j)},y^{ij})} \tilde A^{\alpha j}_+} \P_{\mathcal C_{M \times R}^j}e^{\int_{(x^{ij(j)},y^{ij})} \tilde A^{\alpha j}_+}
\end{equation}
We have moreover
\begin{eqnarray}
\P_{\mathcal C_{M \times R}^j}e^{\int_{(\hat x^{ij(j)},\hat y^{ij})}^{( x^{ij(j)},y^{ij})} \tilde A^{\alpha j}_+} & = & \P_{\varphi^{ij}(\mathcal C_{M \times R}^i)}e^{\int_{(\varphi^{ij}(\hat x^{ij(i)}),\hat y^{ij})}^{( \varphi^{ij}(x^{ij(i)}),y^{ij})} \tilde A^{\alpha j}_+} \\
& = & \P_{\mathcal C_{M \times R}^i}e^{\int_{(\hat x^{ij(i)},\hat y^{ij})}^{( x^{ij(i)},y^{ij})} \varphi^{ij*} \tilde A^{\alpha j}_+}
\end{eqnarray}
Since $\varphi^{ij*} \tilde A^{\alpha j}_+ = \tilde g^{(\alpha i)(\alpha j)-1}_{++} \tilde A^{\alpha i}_+ \tilde g^{(\alpha i)(\alpha j)}_{++} + \tilde g^{(\alpha i)(\alpha j)-1}_{++}d_{M \times R}\tilde g^{(\alpha i)(\alpha j)}_{++}$ we have
\begin{equation}
 \P_{\mathcal C_{M \times R}^i}e^{\int_{(\hat x^{ij(i)},\hat y^{ij})}^{( x^{ij(i)},y^{ij})} \varphi^{ij*} \tilde A^{\alpha j}_+} = \tilde g^{(\alpha i)(\alpha j)}_{++}(\hat x^{ij(i)},\hat y^{ij})^{-1} \P_{\mathcal C_{M \times R}^i}e^{\int_{(\hat x^{ij(i)},\hat y^{ij})}^{( x^{ij(i)},y^{ij})} \tilde A^{\alpha i}_+} \tilde g^{(\alpha i)(\alpha j)}_{++}(x^{ij(i)},y^{ij})
\end{equation}
We see then that
\begin{eqnarray}
& & \P_{\mathcal C_{M \times R}^i} e^{\int^{(\hat x^{ij(i)},\hat y^{ij})} \tilde A^{\alpha i}_+} \tilde g^{(\alpha i)(\alpha j)}_{++}(\hat x^{ij(i)},\hat y^{ij}) \P_{\mathcal C_{M \times R}^j}e^{\int_{(\hat x^{ij(j)},\hat y^{ij})} \tilde A^{\alpha j}_+} \nonumber \\
& & \quad = \P_{\mathcal C_{M \times R}^i} e^{\int^{(x^{ij(i)},y^{ij})} \tilde A^{\alpha i}_+} \tilde g^{(\alpha i)(\alpha j)}_{++}(x^{ij(i)},y^{ij}) \P_{\mathcal C_{M \times R}^j}e^{\int_{(x^{ij(j)},y^{ij})} \tilde A^{\alpha j}_+}
\end{eqnarray}
The formula is then well defined since it is independent of the arbitrary choice of points $\{y^{ij}\}_{ij}$. We can then defined the composite holonomy by
\begin{eqnarray}
\hol_{\tilde A_+}(\{\mathcal C^i_{M \times R}\}_i) & = & \P_{\mathcal C_R}e^{\int_{y_0}^{y^{ij}} \int_{x_0}^{\chi_S^ih(y)} T^{\alpha i}(x,y)D^\alpha_P \tilde A^{i(\alpha)}_{Qx} T^{\alpha i}(x,y)^{-1}} \P_{\mathcal C_R}e^{\int_{y_0}^{y^{ij}} \tilde A^{i(\alpha)}_{Qx_0}} \P_{\mathcal C_M^i}e^{\int_{x_0}^{x^{ij(i)}} A^\alpha_P} \tilde g^{(\alpha i)(\alpha j)}_{++}(x^{ij(i)},y^{ij}) \nonumber \\
& & \times \P_{\mathcal C_R}e^{\int_{y^{ij}}^{y^{jk}} \int_{x^{ij(j)}}^{\chi^j_Sh(y)} T^{\alpha j}(x,y)D^\alpha_P \tilde A^{j(\alpha)}_{Qx} T^{\alpha j}(x,y)^{-1}} \P_{\mathcal C_R}e^{\int_{y^{ij}}^{y^{jk}} \tilde A^{j(\alpha)}_{Qx^{ij(j)}}} \P_{\mathcal C_M^j}e^{\int_{x^{ij(j)}}^{x^{jk(j)}} A^\alpha_P} \tilde g^{(\alpha j)(\alpha k)}_{++}(x^{jk(j)},y^{jk}) \nonumber \\
& & ... \nonumber \\
& & \times \P_{\mathcal C_R}e^{\int_{y^{zi}}^{y_0} \int_{x^{zi(z)}}^{\chi^z_Sh(y)} T^{\alpha z}(x,y) D^\alpha_P \tilde A^{z(\alpha)}_{Qx} T^{zi}(x,y)^{-1}} \P_{\mathcal C_R}e^{\int_{y^{zi}}^{y_0} \tilde A^{z(\alpha)}_{Qx^{zi(i)}}} \P_{\mathcal C_M^i}e^{\int_{x^{zi(i)}}^{x_0} A^\alpha_P} \label{holocomp2}
\end{eqnarray}
where $V^i, V^j, ..., V^z$ are the charts crossed by $\mathcal C_R$ (we have supposed that $\forall i$, $\mathcal C_M^i \subset U^\alpha$). If we relax both the two assumptions then the composite holonomy formula is a mixing of the formulea (\ref{holocomp1}) and (\ref{holocomp2}).

\section{Applications}
In this section we apply the formalism of the principal composite bundle to model two physical problems : 
\begin{itemize} 
\item the dynamics of a quantum system interacting with a classical environment and described by an active space and non-abelian geometric phases,
\item the transport of a classical Dirac spinor field coupled with the gravitational field associated with the curved space-time of the general relativity.
\end{itemize}
In this two situations we present the application of the theorem 1.
\subsection{Non-abelian geometric phases of a quantum system interacting with a classical environment}
We consider a quantum system described by the Hilbert space $\mathcal H$ and the free self-adjoint Hamiltonian $H_0 \in \mathcal L(\mathcal H)$ ($\mathcal L(\mathcal H)$ denotes the space of linear operators of $\mathcal H$). The quantum system interacts with an environment described by $n$ classical parameters $\vec R$ via the self-adjoint interaction operator $\vec R \mapsto H_I(\vec R) \in \mathcal L(\mathcal H)$ (we suppose that $H_0$ and the family $\{H_I(\vec R)\}_{\vec R \in \mathbb R^n}$ have a common domain in which each operator is restricted). The total Hamiltonian $H(\vec R) = H_0 + H_I(\vec R)$ describes the quantum system interacting with its environment. If we suppose that there exists $m$ smooth constraints on the classical parameters $\vec R$ : $\{f^l(\vec R) = 0\}_{l=1,...,m}$, $f^l : \mathbb R^n \to \mathbb R$, then the admissible $\vec R$ form a smooth $(n-m)$-dimensional submanifold $M$ of $\mathbb R^n$. These constraints can have for origin the fact that the parameters could be not physically independent or else they could be chosen by an experimentalist who controls the environment. We denote by $(x^\mu)_{\mu = 1,...,n-m}$ a coordinate system on $M$ (and we write $\vec R = x$ if $\vec R \in M$). Let $[0,T]$ be the fixed time interval of the evolution ($0$ can be the date of the preparation of the system and $T$ can be the date of the experimental measurement). We consider only closed evolutions $[0,T] \ni t \mapsto \vec R(t) \in M$ such that $\vec R(0) = \vec R(T) = x_0$ with $H_I(x_0) = 0$ ($x_0$ corresponds to the off interaction). Since $\vec R(0)=\vec R(T)$ for all evolutions, we can consider $t \in [0,T]$ as a coordinate on the circle $S^1$.\\
We suppose that there exist $N < \dim \mathcal H$ orthonormalized $x$-dependent vectors $(|\mathsf A,x \rangle^\alpha \in \mathcal H)_{\mathsf A=1,...,N}$ (depending on a system of local charts $\{U^\alpha\}_\alpha$ on $M$) such that
\begin{equation}
\label{adiabcond}
\forall t \in [0,T], \quad U(t,0) P(x_0) = P(x(t)) U(t,0) \qquad \text{for all evolution $t \mapsto x(t)$}
\end{equation}
where $P(x) = \sum_{\mathsf A=1}^N |\mathsf A,x \rangle^\alpha {^\alpha}\langle \mathsf A,x |$ is the rank $N$ orthogonal projector on the space spaned by $(|x,\mathsf A \rangle^\alpha \in \mathcal H)_{\mathsf A=1,...,N}$ (the active space), and where $U(t,0)$ is the evolution operator, i.e.
\begin{equation}
\ihbar \frac{dU(t,0)}{dt} = H(x(t)) U(t,0) \qquad U(t,0) = \Te^{- \ihbar^{-1} \int_0^t H(x(t'))dt'}
\end{equation}
$\T$ is the time ordering operator. $\forall x \in U^\alpha \cap U^\beta$, $g^{\alpha \beta}(x)_{\mathsf{AB}} = {^\alpha}\langle \mathsf A,x|\mathsf B,x\rangle^\alpha$ is an element of a basis change matrix within the active space. Such a set of vectors is for example given by an adiabatic theorem (see for example ref. \cite{nenciu}) (in that case each $|\mathsf A,x \rangle$ is an eigenvector of $H(x) = H_0+H_I(x)$). We suppose that the initial wave function of the quantum system is $\psi(0) = |\mathsf A,x_0 \rangle^\alpha$; under the assumption eqn.(\ref{adiabcond}) we can prove (see ref. \cite{viennot,viennot2}) that
\begin{equation}
\label{wavefunct}
\psi(T) = \sum_{\mathsf B=1}^N \left[\Te^{- \ihbar^{-1} \int_0^T E^\alpha(x(t))dt - \int_0^T A_{P\mu}^\alpha(x(t)) \frac{dx^\mu(t)}{dt} dt} \right]_{\mathsf{BA}} |\mathsf B,x_0 \rangle^\alpha
\end{equation}
with
\begin{equation}
E^\alpha(x)_{\mathsf{AB}} = {^\alpha}\langle \mathsf A,x|H(x)|\mathsf B,x \rangle^{\alpha} \qquad A_{P\mu}^\alpha(x)_{\mathsf{AB}} = {^\alpha}\langle \mathsf A,x|\frac{\partial}{\partial x^\mu}|\mathsf B,x \rangle^\alpha
\end{equation}
We have moreover supposed that $\mathcal C_M$ (the path parametrized by $t\mapsto x(t)$) is totally included in $U^\alpha$.\\
The quantum dynamics is then described by the composite bundle $P_+(M \times S^1,U(N)) \to M \times S^1 \to S^1$ (where $U(N)$ is the Lie group of the order $N$ unitary matrices). The structure bundle $P(M,U(N)) \to M$ is defined by the transition function $g^{\alpha \beta} \in \Omega^0(U^\alpha \cap U^\beta,U(M))$. Since the base bundle $M \times S^1 \to S^1$ is trivial, the composite transition functions are simply $g_{++}^{(\alpha i)(\beta j)}(x,t) = g^{\alpha \beta}(x)$ for all opens $V^i$ and $V^j$ of $S^1$, and the torsion functions are reduced to the identity map. The manifold $P_+$ is then $P \times S^1$. The transversal bundles are then the trivial bundles $Q_x^\alpha = S^1 \times \phi^\alpha_{Px}(G) \to S^1$ ($\phi^\alpha_{Px}(G) \simeq \pi^{-1}_P(x)$). The composite bundle is endowed with the composite connection defined by the gauge potential
\begin{equation}
\tilde A_+^\alpha = \ihbar^{-1}E^\alpha(x)dt + A_{P\mu}^\alpha(x)dx^\mu \in \Omega^1(U^\alpha \times S^1,\mathfrak u(N))
\end{equation}
where $A_P \in \Omega^1(U^\alpha,\mathfrak u(N))$ is the gauge potential of $P(M,U(N))$ and where $\tilde A_{Qx}^{(\alpha)}(t) = \ihbar^{-1}E^\alpha(x)dt \in \Omega^1(S^1,\mathfrak u(N))$ is the gauge potential of $Q_x^\alpha(S^1,U(N))$ ($\mathfrak u(N)$ is the Lie algebra of the order $N$ anti-self-adjoint matrices). We note that the gauge of decomposition is conserved while the basis changes of the active space are assumed to be time-independent. The non-abelian phase appearing in the expression of the wave function eqn.(\ref{wavefunct}) is the holonomy of the section $h \in \Gamma(S^1,M \times S^1)$ defined by $h(t)=(x(t),t)$:
\begin{equation}
\Te^{\ihbar^{-1} \int_0^T E^\alpha(x(t))dt + \int_0^T A_{P\mu}^\alpha(x(t)) \frac{dx^\mu(t)}{dt} dt} = \hol_{\tilde A_+}(\mathcal C_{M \times S^1})
\end{equation}

The study of quantum dynamics by the non-abelian geometric phase formulation is well establish when $\forall x,x' \in M$, $\forall \mu$, $[E(x),A_\mu(x')]=0$ (a such assumption is satisfied when all vectors $|\mathsf A,x \rangle$ are eigenvectors associated with a single $N$ degenerate eigenvalue $e(x)$ of $H(x)$). In that case we have
\begin{equation}
\hol_{\tilde A_+}(\mathcal C_{M \times S^1}) = \Te^{\ihbar^{-1} \int_0^T E^\alpha(x(t))dt} \P_{\mathcal C_M}e^{\oint A_P^\alpha}
\end{equation}
The separation of the dynamical phase and of the geometric phase permits the use of the properties of each one to understand the dynamics. In the case of an $N$ fold degenerate eigenvalue, $\Te^{\ihbar^{-1} \int_0^T E^\alpha(x(t))dt} = e^{\ihbar^{-1} \int_0^T e(x(t))dt} id_N$ is just an abelian phase ($id_N$ is the order $N$ identity matrix), the properties concerning the state transitions are then encoded only by the geometry of $P(M,G)$. This fact is used to develop quantum control methods (see ref. \cite{control}). However, if the generators do not commute, $[E(x),A_\mu(x')] \not=0$, the usual formulae does not separate the geometric and the dynamical phase. By applying the theorem 1, we find another separation if
\begin{equation}
\frac{\partial}{\partial x^\mu} E^\alpha(x) +[A_{P\mu}^\alpha(x),E^\alpha(x)] = 0 \qquad \forall x \in U^\alpha, \forall \mu
\end{equation}
In other words, in place of assuming the non-local commutation of the dynamical and the geometric generators, we assume that the dynamical generator is a local geometric coinvariant (by local we mean that the condition depends only on one point, and a geometric coinvariant is the analogue of a dynamical invariant where the gauge potential takes the place of the Hamiltonian). We have then
\begin{equation}
\hol_{\tilde A_+^\alpha}(\mathcal C_{M \times S^1}) = e^{\ihbar^{-1} E^\alpha(x_0) T} \P_{\mathcal C_M}e^{\oint A^\alpha_P}
\end{equation}
However the intertwining curvature vanishing condition can be very drastic. We can relax it, by assuming the following condition
\begin{equation}
\label{cond2}
\frac{\partial}{\partial x^\mu} E^\alpha(x) + [A^\alpha_\mu(x),E^\alpha(x)] = \lambda_\mu^\alpha(x) id_N \qquad \forall x,\forall \mu
\end{equation}
where $\lambda_\mu$ is a real smooth function. In that case, we have
\begin{equation}
\hol_{\tilde A_+^\alpha}(\mathcal C_{M \times S^1}) = e^{\ihbar^{-1} \int_0^T \int_{x_0}^{x(t)} \lambda_\mu^\alpha(x) dx^\mu dt} e^{\ihbar^{-1} E^\alpha(x_0) T} \P_{\mathcal C_M}e^{\oint A^\alpha_P}
\end{equation}
The intertwing term $e^{\ihbar^{-1} \int_0^T \int_{x_0}^{x(t)} \lambda_\mu^\alpha(x) dx^\mu dt}$ is just an abelian phase which does not participate in the transitions between the states $\{ |\mathsf A,x \rangle^\alpha \}_{\mathsf A}$.\\
By using the Leibniz rule, we have a non-abelian generalization of the Hellmann-Feynman theorem
\begin{equation}
\frac{\partial}{\partial x^\mu} E^\alpha_{\mathsf A \mathsf B} = -[A^\alpha_\mu,E^\alpha]_{\mathsf A \mathsf B} + {^\alpha}\langle \mathsf A,x| \frac{\partial H}{\partial x^\mu} | \mathsf B, x \rangle^\alpha
\end{equation}
The assumption (\ref{cond2}) is then realized if $\{|\mathsf A,x \rangle^\alpha \}_{\mathsf A}$ is a set of eigenvectors of all operators $\frac{\partial H_I}{\partial x^\mu}$ associated with single $N$ degenerate eigenvalues:
\begin{equation}
\frac{\partial H_I}{\partial x^\mu} |\mathsf A,x \rangle^\alpha = \lambda_\mu^\alpha(x) |\mathsf A, x \rangle^\alpha
\end{equation}

Remark : even if $Q^\alpha_x \to S^1$ is trivial the holonomy in this bundle does not vanish since the base manifold $S^1$ is topologicaly untrivial.

\subsection{Dirac spinor field transport in a curved spacetime}
\noindent \textit{In this section we suppose that $\hbar = c = e = 1$}\\

Let $M$ be the spacetime manifold which is endowed with a metric $g_{\mu \nu}(x)$ and with a linear connection (the Christoffel symbols) $\Gamma^{\mu}_{\rho \nu}(x)$ ($\{x^\mu\}_{\mu=0,...,3}$ is a spacetime coordinates system). To describe spinor fields in the curved spacetime, we need to consider two other entities. The first one is a Lorentz connection (so-called a spin-connection) $\omega = \omega^{\mathsf A \mathsf B}_\mu(x) M_{\mathsf A \mathsf B} dx^\mu \in \Omega^1(M,\mathfrak h)$ where $\{M_{\mathsf A \mathsf B}\}_{\mathsf A,\mathsf B=0,...,3}$ are the generators of the Lie algebra $\mathfrak h$ of the Lorentz group $H = SO(3,1)$. The second one is a set of vectors forming an orthogonal tangent basis and called tetrads or vierbeins, $\{e^{\mathsf A}_{\mu}(x)\}_{\mathsf A,\mu=0,...,3}$. A vierbein is transformed under a local Lorentz transformation $h(x) \in H$ as $e^{\mathsf A \prime}_\mu(x) = h(x)^{\mathsf A}_{\mathsf B} e^{\mathsf B}_\mu(x)$ ($h(x)^{\mathsf A}_{\mathsf B}$ is the $(\mathsf A,\mathsf B)$-matrix element of $h(x)$). The vierbeins can be interpreted as the gravitational field. The two representations of the gravity are related by the following equations:
\begin{equation}
g_{\mu \nu}(x) = e^{\mathsf A}_{\mu}(x) e^{\mathsf B}_{\nu}(x) \eta_{\mathsf A \mathsf B} = e^{\mathsf A}_\mu(x) e_{\mathsf A \nu}(x)
\end{equation}
where $\eta$ is the Minkowski metric and where $e_{\mathsf A \nu} = \eta_{\mathsf A \mathsf B} e^{\mathsf B}_\nu$,
\begin{equation}
\Gamma^\mu_{\rho \nu}(x) = e^\mu_{\mathsf A}(x) \partial_\rho e^{\mathsf A}_\nu(x) + e^\mu_{\mathsf A}(x) \omega^{\mathsf A \mathsf B}_\rho(x) e_{\mathsf B \nu}(x)
\end{equation}
where $(e^\mu_{\mathsf A})_{\mu, \mathsf A}$ is the formal inverse (in the matrix sense) of $(e_\mu^{\mathsf A})_{\mu, \mathsf A}$ : $e^\mu_{\mathsf A} e^{\mathsf A}_\nu = \delta^{\mu}_\nu$ and $e^{\mu}_{\mathsf A} e^{\mathsf B}_\mu = \delta_{\mathsf A}^{\mathsf B}$ ($\delta^\mu_\nu$ is the Kronecker symbol). Conversely we have
\begin{equation}
\omega^{\mathsf A \mathsf B}_\rho(x) =  e^{\mathsf A}_\mu(x) \partial_\rho e^{\mathsf B \mu}(x) +  e^{\mathsf A}_\mu(x) \Gamma^{\mu}_{\rho \nu}(x) e^{\mathsf B \nu}(x)
\end{equation}

A Dirac spinor field $\psi$ of mass $m$ in the curved spacetime satisfies the Einstein-Dirac equation (ref. \cite{brill})
\begin{equation}
(\imath \gamma^{\mathsf A} e_{\mathsf A}^\mu(x) \nabla_\mu -m) \psi(x) = 0
\end{equation}
where $\{\gamma^{\mathsf A}\}_{\mathsf A=0,...,3}$ are the Dirac matrices and where $\nabla_\mu$ is the spinorial covariant derivative defined by
\begin{equation}
\label{nabla}
\nabla_\mu  =  \frac{\partial}{\partial x^\mu} + \omega^{\mathsf A \mathsf B}_\mu(x) \mathfrak D(M_{\mathsf A \mathsf B})
\end{equation}
where $\mathfrak D$ is $(1/2,0)\oplus(0,1/2)$ representation of the Lorentz group $H$ (we denotes by the same symbol the induced representation of its Lie algebra), i.e.
\begin{equation}
\mathfrak D(M_{\mathsf A \mathsf B}) = \frac{1}{4} \left[ \gamma_{\mathsf A}, \gamma_{\mathsf B} \right]
\end{equation}
Following Sardanashvily ref. \cite{sardana2,sardana3} the gauge theory is described by a composite bundle. Let $G = GL(4,\mathbb R)$ be the group of invertible order $4$ matrices. Let $P_+(M,G) \to M$ be the principal $G$-bundle of the tangent frames of $M$. For a good open cover $\{V^i\}_i$ of $M$, let $\phi^i_{TF}:V^i \times G \to P_+$ be the local trivialisation of this bundle. Viewed as a (fixed) matrix $e$ with elements $e_\mu^{\mathsf A}$, the vierbeins belong to $G$ ($[e^{-1}]^\mu_{\mathsf A} =  e_{\mathsf A}^\mu$). The equivalence class of the vierbeins under the constant Lorentz transformations, $eH = \{e_\mu^{\mathsf A} h^\mu_\nu , h \in H \}$, belongs to $G/H$. We must then view the group $G$ as a principal $H$-bundle, $G(G/H,H) \to G/H$, with local trivialisation $\phi^\alpha_G : U^\alpha \times H \to G$ for a good open cover $\{U^\alpha\}_\alpha$ of the manifold $G/H$. It is then natural to consider the manifold $S = P_+ / H$ which has the fiber bundle structure $S(M,G/H) \to M$. We have then the following principal composite $G(G/H,H)$-bundle $P_+(S,H) \to S(M,G/H) \to M$ where the map $\pi_{+} : P_+ \to S$ is the canonical projection associated with the quotient $P_+/H$. The diffeomorphism $\chi^i_S : \pi^{-1}_S(V^i) \to V^i \times G/H$ is just the map induced by $\phi^{i -1}_{TF} : P_+ \to V^i \times G$. The total twisted bundle $\{P_+^i(G/H \times V^i, H) \to G/H \times V^i\}_i$ is then defined by the local trivialisation $\phi^{\alpha i}_{++}: \begin{array}{rcl} U^\alpha \times V^i \times H & \to & P_+^i \\ (eH,x,h) & \mapsto & \phi^i_{TF}(x,\phi^\alpha_G(eH,h)) \end{array}$.\\
This composite bundle permits consideration of the Lorentz connection as a composite connection. We endow the structure bundle $G(G/H),H) \to G/H$ by the Cartan connection associated with the following gauge potential
\begin{equation}
A_{G}(eH) =  e_\mu^{\mathsf A} de^{\mathsf B \mu} M_{\mathsf A \mathsf B} \in \Omega^1(G/H,\mathfrak h)
\end{equation}
and the transversal bundle $Q_{eH}(M,H) \to M$ by the spinorial representation of the linear connection which is associated with the following gauge potential
\begin{equation}
\tilde A_{QeH}(x) =  e_\mu^{\mathsf A} \Gamma^\mu_{\rho \nu}(x) e^{\mathsf B \nu} M_{\mathsf A \mathsf B} dx^\rho \in \Omega^1(M,\mathfrak h)
\end{equation}
(We omit the chart indices, which play no role in this discussion). Let $ \sigma = (x \mapsto e(x)H) \in \Gamma(M,S)$ be a local section of $S(M,G/H) \to M$. We have
\begin{eqnarray}
\sigma^* \tilde A_+(x) & = & \left(e_\mu^{\mathsf A}(x) \partial_\rho e^{\mathsf B \mu}(x) + e_\mu^{\mathsf A}(x) \Gamma^\mu_{\rho \nu}(x) e^{\mathsf B \nu}(x) M_{\mathsf A \mathsf B} \right)dx^\rho \\
& = & \omega
\end{eqnarray}
The Lorentz gauge potential is then the composition of the Cartan gauge potential and of the linear potential. $\mathfrak D(\omega)$ is the gauge potential associated with the spinorial covariant derivative $\nabla_\mu$ eqn.(\ref{nabla}). In constrast to the approach of Tresguerres ref. \cite{tresguerres} in which the vierbeins appear as a translational gauge potential of a Poincar\'e gauge theory of the gravity, in this approach the vierbeins are not fixed by the connection whereas their equivalence classes form the auxiliary spacetime $G/H$ on which the gauge theory is built.\\
Holonomies associated with the Lorentz connection play an important role in the quantization of the gravity (see ref. \cite{rovelli, thiemann}). Let two separate particles at the spacetime point $x_0$ with the same spinor state $\hat \psi_0$. The first particle is transported along the worldline $\mathcal C_a$ from $x_0$ to $x_1$ and the second particle is transported along another worldline $\mathcal C_b$ from $x_0$ to $x_1$. This situation means that the particles are described by the semi-classical spinor fields as $\psi_a(x) = \int \hat \psi_a(s) \delta(x-x_a(s))ds$ where $s \mapsto x_a(s)$ is the parametrisation of the worldline of the particle $a$, and $\hat \psi_a(s)$ is the spinor state at the proper time $s$ along the worldline with $\hat \psi_a(s_0) = \hat \psi_0$ ($x_a(s_0)=x_0$). After the transportations ($x_a(s_{1a})=x_b(s_{1b})=x_1$) the spinor states of the two particles are related by
\begin{equation}
\hat \psi_{a}(s_{1a}) = \mathfrak D(\hol_{\tilde A_+}(\mathcal C_{M \times G/H})) \hat \psi_{b}(s_{1b})
\end{equation}
where $\mathcal C_M = \mathcal C_a \circ \mathcal C_b^{-1} \in \mathcal L_{x_1}(M)$ and $\mathcal C_{M \times G/H} = \{(x(s),e(x(s))H), s \in [0,s_{1a}+s_{2b}-2s_0] \} \in \mathcal L_{(x_1,e(x_1)H)}(M \times G/H)$, $s$ being the curvilinear coordinate along $\mathcal C_M$ and $x \mapsto e(x)H$ being a local section of $S(M,G/H) \to M$ over $\mathcal C_M$ (associated with the vierbeins $x \mapsto e^{\mathsf A}_\mu(x)$). By applying the theorem 1 we have
\begin{eqnarray}
\P_{\mathcal C_{M }} e^{\imath \oint \omega} & = & \hol_{\tilde A_+}(\mathcal C_{M \times G/H}) \\
& = & \P_{\mathcal C_M}e^{\oint \int_{e(x_1)H}^{e(x)H} T(eH,x) D_G\tilde A_{QeH\mu} T(eH,x)^{-1} dx^\mu}  \hol_{\tilde A_{Qe(x_1)H}}(\mathcal C_M) \hol_{A_G}(\mathcal C_{G/H})
\end{eqnarray}
where $\hol_{\tilde A_{Qe(x_1)H}}(\mathcal C_M)$ is the holonomy associated with the representation of the linear connection on the spinor states at the base point $x_1$ and where $\hol_{A_G}(\mathcal C_{G/H})$ is the holonomy associated with the Cartan connection. The intertwining curvature is (see the appenix)
\begin{equation}
D_G \tilde A_{QeH} = \left(e^{\mathsf B}_\mu de^{\mathsf A}_\lambda g^{\lambda \nu} - e^{\mathsf B \nu} de^{\mathsf A \lambda} g_{\lambda \mu} \right) \wedge \Gamma^\mu_{\rho \nu} dx^\rho M_{\mathsf{AB}}
\end{equation}

 We can note an interesting analogy between the non-abelian geometric phases treated in the previous application: the Cartan connection $A_{G}(eH)$ plays the role of the geometric phase generator, $\tilde A_{QeH}$ plays the role of a dynamical phase generator, the vierbeins play the role of the active space basis vectors and the Christoffel symbol $\Gamma^\mu_{\rho \nu}$ takes the place of the Hamiltonian.  

\subsection*{Acknowledgments}
The author thanks Professor John P. Killingbeck for his help.

\appendix
\section{Demonstration of theorem 1}
Since all relevant quantities are defined on the single local chart $U^\alpha \times V^i$, we omit the indices $\alpha$ and $i$.
\begin{equation}
\P_{\mathcal C_{M \times R}}e^{\oint \tilde A_+} = \P_{\mathcal C_R}e^{\oint \left(\tilde A_{Q \chi_S h(y)}(y) + (h^* \chi_S^* A_P)(y) \right)}
\end{equation}
We split this expression by using the intermediate representation theorem (see ref. \cite{messiah})
\begin{equation}
\P_{\mathcal C_{M \times R}}e^{\oint \tilde A_+} = \P_{\mathcal C_R}e^{\oint \P_{\mathcal C_M}e^{\int_{x_0}^{\chi_Sh(y)} A_P} \tilde A_{Q \chi_S h(y)} \P_{\mathcal C_M}e^{- \int_{x_0}^{\chi_S h(y)} A_P}} \P_{\mathcal C_M}e^{\oint A_P}
\end{equation}
Moreover we have
\begin{equation}
\P_{\mathcal C_M} e^{\int_{x_0}^x A_P} \tilde A_{Q x} \P_{\mathcal C_M}e^{- \int_{x_0}^x A_P} - \tilde A_{Qx_0} = \int_{x_0}^x d_M \left(\P_{\mathcal C_M} e^{\int_{x_0}^x A_P} \tilde A_{Q x} \P_{\mathcal C_M}e^{- \int_{x_0}^x A_P} \right)
\end{equation}
By using the Leibniz rule with $d_M$ in the r.h.s. of the preceding equation, and the fact that $d_M \P_{\mathcal C_M}e^{\int_{x_0}^x A_P} = \P_{\mathcal C_M}e^{\int_{x_0}^x A_P} A_P$ and $d_M \P_{\mathcal C_M}e^{- \int_{x_0}^x A_P} = - A_P \P_{\mathcal C_M}e^{- \int_{x_0}^x A_P}$, we find that
\begin{eqnarray}
\P_{\mathcal C_M} e^{\int_{x_0}^x A_P} \tilde A_{Q x} \P_{\mathcal C_M}e^{- \int_{x_0}^x A_P} - \tilde A_{Qx_0} & = & \int_{x_0}^x \P_{\mathcal C_M} e^{\int_{x_0}^x A_P} (d_M \tilde A_{Qx} + [A_P,\tilde A_{Qx}]) \P_{\mathcal C_M} e^{- \int_{x_0}^x A_P} \\
& = & \int_{x_0}^x \P_{\mathcal C_M} e^{\int_{x_0}^x A_P} D_P \tilde A_{Qx} \P_{\mathcal C_M} e^{- \int_{x_0}^x A_P}
\end{eqnarray}
We then have
\begin{equation}
\P_{\mathcal C_{M \times R}}e^{\tilde A_+} = \P_{\mathcal C_R}e^{\oint \left( \int_{x_0}^{\chi_Sh(y)} \P_{\mathcal C_M}e^{\int_{x_0}^x A_P} D_{P\mu} \tilde A_{Qxa} \P_{\mathcal C_M}e^{- \int_{x_0}^x A_P} dx^\mu + \tilde A_{Qx_0 a} \right)dy^a} \P_{\mathcal C_M}e^{\oint A_P}
\end{equation}
Finally by using again the intermediate representation theorem, we have
\begin{equation}
\P_{\mathcal C_{M \times R}}e^{\tilde A_+} = \P_{\mathcal C_R}e^{\oint \P_{\mathcal C_R}e^{\int_{y_0}^y \tilde A_{Qx_0}} \int_{x_0}^{\chi_Sh(y)} \P_{\mathcal C_M}e^{\int_{x_0}^x A_P} D_{P\mu} \tilde A_{Qxa} \P_{\mathcal C_M}e^{- \int_{x_0}^x A_P} dx^\mu \P_{\mathcal C_R}e^{-\int_{y_0}^y \tilde A_{Qx_0}} dy^a} \P_{\mathcal C_R}e^{\int_{y_0}^y \tilde A_{Qx_0}} \P_{\mathcal C_M}e^{\oint A_P}
\end{equation}

\section{Intertwining curvature of gravity}
\begin{equation}
D_G \tilde A_{QeH} = de^{\mathsf A}_\mu \wedge \Gamma^\mu_{\rho \nu} dx^\rho e^{\mathsf B \nu} M_{\mathsf{AB}} +  e^{\mathsf A}_\mu \Gamma^\mu_{\rho \nu} de^{\mathsf B \nu} \wedge dx^\rho M_{\mathsf{AB}} +  e^{\mathsf A}_\mu de^{\mathsf B \mu} \wedge  e^{\mathsf C}_\lambda \Gamma^\lambda_{\rho \nu} dx^\rho e^{\mathsf D \nu} [M_{\mathsf{AB}},M_{\mathsf{CD}}]
\end{equation}
The commutation relations of the Lorentz algebra $\mathfrak h$ being
\begin{equation}
  [M_{\mathsf{AB}},M_{\mathsf{CD}}] = - \eta_{\mathsf{BD}} M_{\mathsf{AC}} + \eta_{\mathsf{BC}} M_{\mathsf{AD}} + \eta_{\mathsf{AD}} M_{\mathsf{BC}} - \eta_{\mathsf{AC}} M_{\mathsf{BD}}
\end{equation}
we have
\begin{eqnarray}
D_G \tilde A_{QeH} & = & de^{\mathsf A}_\mu e^{\mathsf B \nu} \wedge \Gamma^\mu_{\rho \nu} dx^\rho M_{\mathsf{AB}} + e^{\mathsf A}_\mu de^{\mathsf B \nu} \wedge \Gamma^\mu_{\rho \nu} dx^\rho M_{\mathsf{AB}} \\
& & - e^{\mathsf A}_\mu de^{\mathsf B \mu} e^{\mathsf C}_\lambda e^\nu_{\mathsf B} \wedge \Gamma^\lambda_{\rho \nu} dx^\rho M_{\mathsf{AC}} + e^{\mathsf A}_\mu de^{\mathsf B \mu} e_{\mathsf B \lambda} e^{\mathsf D \nu} \wedge \Gamma^\lambda_{\rho \nu} dx^\rho M_{\mathsf{AD}} \\
& & + \underbrace{e^{\mathsf A}_\mu de^{\mathsf B \mu} e^{\mathsf C}_\lambda e^\nu_{\mathsf A}}_{de^{\mathsf B \nu} e^{\mathsf C}_\lambda} \wedge \Gamma^\lambda_{\rho \nu} dx^\rho M_{\mathsf{BC}} - \underbrace{e^{\mathsf A}_\mu de^{\mathsf B \mu} e_{\mathsf A \lambda} e^{\mathsf D \nu}}_{de^{\mathsf B \mu} e^{\mathsf D \nu} g_{\mu \lambda}} \wedge \Gamma^\lambda_{\rho \nu} dx^\rho M_{\mathsf{BD}}
\end{eqnarray}
After the exchange of some double indices we find
\begin{equation}
D_G \tilde A_{QeH} = \left(de^{\mathsf A}_\mu e^{\mathsf B \nu} + e^{\mathsf A}_\mu de^{\mathsf B \nu} - e^{\mathsf A}_\lambda de^{\mathsf C \lambda} e^{\mathsf B}_\mu e^\nu_{\mathsf C} + e^{\mathsf A}_\lambda de^{\mathsf D \lambda} e_{\mathsf D \mu} e^{\mathsf B \nu} + de^{\mathsf A \nu} e^{\mathsf B}_\mu - de^{\mathsf A \lambda} e^{\mathsf B \nu} g_{\lambda \mu} \right) \wedge \Gamma^\mu_{\rho \nu} dx^\rho M_{\mathsf{AB}}
\end{equation}
Since $e^{\mathsf D \lambda} e_{\mathsf D \mu} = \delta^\lambda_\mu \Rightarrow de^{\mathsf D \lambda} e_{\mathsf D \mu} = - e^{\mathsf D \lambda} de_{\mathsf D \mu}$ we have
\begin{equation}
e^{\mathsf A}_\lambda de^{\mathsf D \lambda} e_{\mathsf D \mu} e^{\mathsf B \nu} = - e^{\mathsf A}_\lambda e^{\mathsf D \lambda} de_{\mathsf D \mu} e^{\mathsf B \nu} = - de^{\mathsf A}_\mu e^{\mathsf B \nu}
\end{equation}
Since $e^{\mathsf A}_\lambda e^{\mathsf C \lambda} = \eta^{\mathsf{AC}} \Rightarrow de^{\mathsf A}_\lambda e^{\mathsf C \lambda} = - e^{\mathsf A}_\lambda de^{\mathsf C \lambda}$ we have
\begin{equation}
e^{\mathsf A}_\lambda de^{\mathsf C \lambda} e^\nu_{\mathsf C} e^{\mathsf B}_\mu = - de^{\mathsf A}_\lambda e^{\mathsf C \lambda} e^\nu_{\mathsf C} e^{\mathsf B}_\mu = e^{\mathsf B}_\mu de^{\mathsf A}_\lambda g^{\lambda \nu}
\end{equation}
We then have
\begin{equation}
D_G \tilde A_{QeH} = \left(e^{\mathsf A}_\mu de^{\mathsf B \nu} + de^{\mathsf A \nu} e^{\mathsf B}_\mu +e^{\mathsf B}_\mu de^{\mathsf A}_\lambda g^{\lambda \nu} -e^{\mathsf B \nu} de^{\mathsf A \lambda} g_{\lambda \mu} \right) \wedge \Gamma^\mu_{\rho \nu} dx^\rho M_{\mathsf{AB}}
\end{equation}
Moreover we have
\begin{equation}
e^{\mathsf A}_\mu de^{\mathsf B \nu} M_{\mathsf{AB}} = e^{\mathsf B}_\mu de^{\mathsf A \nu} M_{\mathsf{BA}} = - e^{\mathsf B}_\mu de^{\mathsf A \nu} M_{\mathsf{AB}}
\end{equation}
because $M_{\mathsf{BA}} = - M_{\mathsf{AB}}$. Finally we have
\begin{equation}
D_G \tilde A_{QeH} = \left(e^{\mathsf B}_\mu de^{\mathsf A}_\lambda g^{\lambda \nu} -e^{\mathsf B \nu} de^{\mathsf A \lambda} g_{\lambda \mu} \right) \wedge \Gamma^\mu_{\rho \nu} dx^\rho M_{\mathsf{AB}}
\end{equation}

\end{document}